\documentclass[12pt]{article}
\usepackage{axodraw,a4,epsfig,amssymb}

  \newcommand{\ccaption}[2]{
    \begin{center}
    \parbox{0.85\textwidth}{
      \caption[#1]{\small{#2}}
      }
    \end{center}
    }

\newcommand{\beq}{\begin{equation}}
\newcommand{\eeq}{\end{equation}}
\newcommand{\beqa}{\begin{eqnarray}}
\newcommand{\eeqa}{\end{eqnarray}}

\newcommand{\Eqn}[1]{Eq.~(\ref{#1})}
\newcommand{\Eqns}[2]{Eqs.~(\ref{#1}) and (\ref{#2})}

\newcommand{\MS}{\overline{\rm MS}}
\newcommand{\OS}{{\rm OS}}
\newcommand{\PS}{{\rm PS}}

\newcommand{\cI}{{\cal I}}

\newcommand{\cO}{{\cal O}}

\begin{document}

\pagestyle{empty}
\begin{flushright}
  IPPP/07/41 \\
  DCPT/07/82 \\
\end{flushright}
\vspace*{1cm}
\begin{center}
  {\sc \large 
Combined fixed-order and effective-theory approach\\[5pt] 
to $b\bar{b}$ sum rules } \\
   \vspace*{2cm}
{\bf  Adrian~Signer}\\
\vspace{0.6cm}
{\it Institute for Particle Physics Phenomenology \\
Durham, DH1 3LE, England \\}
  \vspace*{2.8cm} {\bf Abstract} \\ 
\vspace{1\baselineskip}
\parbox{0.9\textwidth}{ We combine the fixed-order evaluation of the
$b\bar{b}$ sum rules with a non-relativistic effective-theory
approach. The combined result for the $n$-th moment includes all terms
suppressed with respect to the leading-order result by ${\cal O}
(\alpha_s^3)$ and ${\cal O} \left((\alpha_s \sqrt{n})^l \alpha_s^2
\right)$, counting $\alpha_s \sqrt{n} \sim 1$. When compared to
experimental data, the moments thus obtained show a remarkable
consistency and allow for an analysis in the whole range $1\le n
\lesssim 16$. }
\end{center}
\vspace*{5mm}
\noindent

\newpage

\setcounter{page}{1}
\pagestyle{plain}


\section{Introduction and outline \label{sec:I}}

Near threshold, the cross section for the production of a $b\bar{b}$
pair, $\sigma(e^+e^-\to b\bar{b})$, is extremely sensitive to the mass
of the bottom quark $m_b$, which allows for a precise determination of
$m_b$.  This is usually done by considering sum
rules~\cite{Novikov:1976tn} and defining the $n$-th moment
\begin{equation}
M_n \equiv \int_0^\infty \frac{ds}{s^{n+1}}\, R_{b\bar{b}}(s)
= \frac{12 \pi^2 e_b^2}{n!} 
\left(\frac{d}{d q^2}\right)^n \Pi(q^2)\big|_{q^2=0}
\label{MnDef}
\end{equation}
where $\Pi(q^2)$ is the vacuum polarization, $e_b=-1/3$ the electric
charge of the bottom quark and $R_{b\bar{b}}(s)\equiv \sigma(e^+e^-\to
b\bar{b})/\sigma(e^+e^-\to \mu^+\mu^-)$ the normalized cross
section. In order to extract $m_b$, the theoretical evaluation of
$M_n$ is compared to the experimental value.  From the experimental
point of view, the moment obtains contributions from the six
$\Upsilon$ bound states and from the continuum cross section above
threshold. For increasing $n$, the contribution from the
experimentally poorly known continuum cross section becomes less and
less relevant due to the suppression $1/s^{n+1}$. As for the choice of
the parameter $n$, there are two complementary approaches. Either $n$
is assumed to be rather small, i.e. $n\lesssim 4$ in which case
$\Pi(q^2)$ is computed in a standard weak coupling fixed-order
approach or $n$ is assumed to be rather large $n\gtrsim 8$ in which
case the moments are evaluated in a non-relativistic effective-theory
approach.

In the standard fixed-order (FO) approach, the vacuum polarization is
written as
\begin{equation}
\Pi(q^2) = \frac{N_c}{(4\pi)^2} \sum_{n\ge 0} \, C_n \, 
\left(\frac{q^2}{4 m_b^2}\right)^n
\label{PiFO}
\end{equation}
where $N_c=3$ is the colour factor and the coefficients $C_n$ are
evaluated as a series in the strong coupling $\alpha_s$. These
coefficients depend on the mass scheme that is used. We will indicate
this dependence by a label $X$, i.e. $m_X$ denotes the bottom quark
mass in a particular scheme and $C_{n, X}$ are the corresponding
coefficients. From the knowledge of $C_{n,X}$, the moments are obtained
as
\begin{equation}
M_{n,X} = \frac{3}{4} N_c\, e_b^2 \,
\frac{1}{(2 m_X)^{2n}}\, C_{n,X}
\label{momFO}
\end{equation}
Being observables, the moments should be scheme independent. However,
since the perturbative series is truncated, there is a residual scheme
dependence left in $M_n$ which again is indicated by the label
$X$. The coefficients $C_n$ have been computed up to
$\cO(\alpha_s^2)$, i.e. three loops for $n\le 8$ in
Ref.~\cite{Chetyrkin:1997mb} and up to $n\le 30$ in
Ref.~\cite{Boughezal:2006uu}. The four-loop coefficient is known for
$n=0,1$~\cite{Chetyrkin:2006xg, Boughezal:2006px} and these results
have been used to obtain precise values for $\overline{m} \equiv
m_{\MS}$, the bottom quark mass in the
$\MS$-scheme~\cite{Boughezal:2006px, Kuhn:2007vp}.

At $l+1$ loops, the coefficients $C_n$ contain terms $n^{-3/2}
(\alpha_s \sqrt{n})^l$.  Thus, if $n$ increases, the higher-order
terms become more important and for $\sqrt{n} \sim \alpha_s^{-1}$ the
standard fixed-order approach completely fails. This is related to the
fact that in a strict expansion in $\alpha_s$ the theory does not
contain bound states. Given that for increasing $n$ the moments are
dominated by the lowest resonances it is thus not surprising that a
FO approach does not very well describe $M_n$ for large
$n$. As a consequence, mass determinations using this
approach~\cite{Kuhn:2001dm,Corcella:2002uu, Boughezal:2006px,
Kuhn:2007vp} use small values of $n$.

In order to describe the weak coupling bound states in the $b\bar{b}$
system, we have to consider the non-relativistic sum rule. The
starting point is the solution to the Schr\"odinger equation
describing a non-relativistic $b\bar{b}$ pair interacting through the
Coulomb potential $- C_F \alpha_s/r$ with $C_F=4/3$ a colour factor.
This resums all terms of the form $v\, (\alpha_s/v)^l$ in
$R_{b\bar{b}}$, where $v$ is the small velocity of the heavy quarks
and, therefore, resums all terms of the form $n^{-3/2} (\alpha_s\,
\sqrt{n})^l$ in $M_n$. Higher order corrections in $\alpha_s$ as well
as $v$ are taken into account using Quantum Mechanics perturbation
theory and counting $\alpha_s \sim v$. This is done most efficiently
in the framework of an effective theory (for a review see
Ref.~\cite{Brambilla:2004jw}). Within the effective-theory (ET)
approach the next-to-next-to-leading order (NNLO) corrections to the
non-relativistic sum rules, including all terms suppressed by
$\alpha_s^2 \sim \alpha_s/\sqrt{n} \sim n^{-1}$ with respect to the
leading-order result, have been computed and used to determine the
bottom quark mass~\cite{BmassNR}.  The theoretical predictions can be
improved upon by resummation of large logarithms~\cite{resumLog} of
the form $(\alpha_s \log v)^l$ and including these terms in the
non-relativistic sum rule leads to a much more robust determination of
the bottom quark mass~\cite{Pineda:2006gx}.

Ultimately the most interesting quantity is $\overline{m}$, the bottom
quark mass in the $\MS$-scheme. Using non-relativistic sum rules,
$\overline{m}$ has to be determined in two steps. First, a so called
threshold mass~\cite{massdef, Beneke:1998rk, Pineda:2001zq} has to be
used which is closely related to the pole mass but accounts for the
cancellation of the renormalon ambiguity in the observable $M_n$. In a
second step, the threshold mass is related to $\overline{m}$. Using
this approach, the moments can be determined reliably for large values
of $n$, as long as non-perturbative contributions are not too
important. However, for small values of $n$ this approach breaks down
due to the neglect of terms suppressed for large $n$ or small $v$. As
an example consider terms of order $\alpha_s^0 v^3$ in $R_{b\bar{b}}$
which are kept at NNLO in an ET approach, whereas terms of order
$\alpha_s^0 v^5$ are dropped. These terms result in contributions of
the order $\alpha^0 n^{-5/2}$ and $\alpha^0 n^{-7/2}$ respectively in
$M_n$. It is clear that the latter are suppressed in the
non-relativistic sum rule, i.e. for large $n$, but their neglect
invalidates the $n\to 1$ limit of the result obtained in the
non-relativistic approach.

Large $n$ and small $n$ applications of the sum rules and the
corresponding determinations of $m_b$ both have their advantages and
disadvantages. From the large $n$ point of view, one advantage is that
due to $M_n \sim 1/(2 m_b)^{2n}$ the moments are much more sensitive
to the bottom quark mass for large $n$. Also, they are virtually
insensitive to the continuum contribution. Since this contribution is
experimentally only known very poorly, small $n$ determinations of the
bottom quark mass crucially rely on the precise treatment of the data
in the threshold region and have to use perturbative QCD input for
$R_{b\bar{b}}$ above threshold. Given that the experimental error in
the $m_b$ determination for small $n$ is dominant, rather subtle
changes in the treatment can have significant effects on the extracted
value of $m_b$ and, in particular, its error. On the other hand, the
perturbative series is much better behaved for small $n$. In fact, the
non-relativistic sum rules suffers from very large corrections and
even though the resummation of the logarithms substantially improves
the behaviour of the perturbative series, the situation is far from
ideal, and the dominant error still comes from the neglect of
higher-order corrections. For completeness we mention again that in
the large $n$ approach $\overline{m}$ cannot be obtained directly but
only through the intermediate use of a threshold mass. However, this
is not necessarily a disadvantage, since threshold masses are
important and useful in their own right.

It is natural to ask whether it is not preferable to combine both
approaches and perform an all $n$ analysis of the sum rule. Since the
two approaches use different techniques and are sensitive to rather
different experimental input it certainly would give increased
credibility if the extracted value of $m_b$ varies very little with
$n$, and it would allow to get a better handle on the determination of
its error. The choice of $n$ in such a combined analysis is only
limited by the non-perturbative corrections. In order to get an
estimate of the importance of these corrections it is useful to
consider the contribution of the gluon condensate to the sum
rule~\cite{Shifman:1978bx, Broadhurst:1994qj}. Even though this
contribution grows rapidly with $n$, for realistic values of the gluon
condesate it is below $0.1\%$ for $n\leq 12$ and reaches about $1\%$
for $n=16$. It is thus legitimate to neglect non-perturbative
corrections to the $b$-quark sum rules as long as $n$ is not chosen to
be too large.

In this paper we consider the first 16 moments, starting in
Section~\ref{sec:II} with the fixed-order approach. Even though the
large-$n$ behaviour of the FO results is better than anticipated, will
find the expected problems for large $n$ and turn in
Section~\ref{sec:III} to the non-relativistic sum rule in order to
illustrate its behaviour as a function of $n$. Finally, in
Section~\ref{sec:IV}, we combine the two approaches by adding to the
non-relativistic sum rules all terms order $\alpha_s^3$ that have been
missed. As we will see this allows to obtain a consistent description
of $M_n$ for $1\le n \le 16$. We refrain from presenting another
extraction of $m_b$, since all theory input used in this analysis has
already been used for a bottom mass
determination~\cite{Boughezal:2006px, Kuhn:2007vp, Pineda:2006gx}. The
main aim of the paper is to establish the fact that a future analysis,
once further improved theoretical results and hopefully better
experimental data is available, should consider the full range of $n$
in order to get a better control of the different systematic
uncertainties.

\section{Fixed order results, small $n$ \label{sec:II}}

In a fixed-order approach, $M_{n, X}$ can be written as in \Eqn{momFO}
and the coefficient $C_{n, X}$ has the structure
\begin{eqnarray}
\label{Cstructure}
C_{n,X} &=& C_{n,X}^{(0)} + \frac{\alpha_s}{\pi}\, C_{n,X}^{(1)} 
+ \frac{\alpha_s^2}{\pi^2}\, C_{n,X}^{(2)} 
+ \frac{\alpha_s^3}{\pi^3}\, C_{n,X}^{(3)} + \ldots \\
&=& C_{n,X}^{(0)} + 
\frac{\alpha_s}{\pi} \left(C_{n,X}^{(10)} + C_{n,X}^{(11)} L_X \right)
+ \frac{\alpha_s^2}{\pi^2} \left( C_{n,X}^{(20)} + 
   C_{n,X}^{(21)} L_X  +    C_{n,X}^{(22)} L_X^2 \right)
\nonumber \\
&& + \ 
\frac{\alpha_s^3}{\pi^3} \left( C_{n,X}^{(30)} + 
   C_{n,X}^{(31)} L_X  + C_{n,X}^{(32)} L_X^2 + 
   C_{n,X}^{(33)} L_X^3 \right) + \ldots
\nonumber
\end{eqnarray}
where we have introduced 
\begin{equation}
L_X \equiv \log\frac{\mu^2}{m_X^2}
\label{logdef}
\end{equation}
In the on-shell scheme we use the notation $C_{n}^{(kl)}\equiv
C_{n,\OS}^{(kl)}$ and we have $C_{n}^{(11)} = C_{n}^{(22)} =
C_{n}^{(33)} = 0$. The logarithmic coefficients of order $\alpha_s^k$,
i.e. $C_{n,X}^{(kl)}$ with $l\ge 1$, can be predicted from the lower
order coefficients of order $\alpha_s^m, \ m<k$. The three-loop
coefficients $C_{n,X}^{(20)}$ have been computed up to $n=8$ in
Ref.~\cite{Chetyrkin:1997mb} and later up to $n=30$ in
Ref.~\cite{Boughezal:2006uu}. The four-loop coefficient
$C_{1,X}^{(30)}$ is also known \cite{Chetyrkin:2006xg,
Boughezal:2006px} but for $n>1$ these coefficients have not yet been
computed.

The relation between $C_{n,X}^{(kl)}$ and $C_{n}^{(kl)}$, the
coefficients in the scheme $X$ and the on-shell scheme respectively,
where the corresponding masses (we denote the pole mass by $m\equiv
m_{\OS}$) are related by
\begin{equation}
m = m_X \left(1 + \frac{\alpha_s}{\pi}\, \delta m^{(1)}_{X} 
 + \frac{\alpha_s^2}{\pi^2}\, \delta m^{(2)}_{X} 
 + \frac{\alpha_s^3}{\pi^3}\, \delta m^{(3)}_{X} +\ldots \right)
\label{massrel}
\end{equation}
is given by
\begin{eqnarray}
C_{n,X}^{(0)} &=& C_{n}^{(0)} \\
C_{n,X}^{(1)} &=& 
   C_{n}^{(10)}
   - 2 n\, C_{n}^{(0)}\, \delta m^{(1)}_{X} \\ 
C_{n,X}^{(2)} &=& 
   C_{n}^{(20)} + C_{n}^{(21)} L_X 
   - 2n \, C_{n}^{(10)}\, \delta m^{(1)}_{X}  \\
  &&  +\  \left[ n(1+2n) (\delta m^{(1)}_{X})^2
       - 2n\, \delta m^{(2)}_{X}
       \right]  C_{n}^{(0)} \nonumber \\
C_{n,X}^{(3)} &=& C_{n}^{(30)} + C_{n}^{(31)} L_X 
   + C_{n}^{(32)}  L_X^2 
    -\ 2n\, \delta m^{(1)}_{X}\, 
          \left[ C_{n}^{(20)} + C_{n}^{(21)} L_X \right] \\
&&  -\  2\, C_{n}^{(21)}\, \delta m^{(1)}_{X} 
   + \left[ n(1+2n) (\delta m^{(1)}_{X})^2 -
           2n\,\delta m^{(2)}_{X}\right]  C_{n}^{(10)} 
\nonumber \\
&& -\ \frac{2n}{3} \left[(1+n)(1+2n) (\delta m^{(1)}_{X})^3
        - 3(1+2n)\, \delta m^{(1)}_{X}\, \delta m^{(2)}_{X}
   + 3\, \delta m^{(3)}_{X}\right] C_{n}^{(0)} \nonumber
\end{eqnarray}
It is clear that theses relations break down for large $n$, due to the
terms of order $(n \alpha_s)^k$. In fact, the behaviour for
$n\to\infty$ seems to be even worse than $n^{-3/2} (\alpha_s
\sqrt{n})^l$ as mentioned in the introduction. This is due to the
shift in the mass scheme. To be precise, $C_{n}$, the coefficients in
the on-shell scheme behave like $n^{-3/2} (\alpha_s \sqrt{n})^l$ for
$n\to \infty$. For any threshold scheme $X$, the factors $\delta
m_{X}^{(l)}$ have to have an additional suppression $\delta
m_{X}^{(l)} \sim \alpha_s$ in order not to destroy the behaviour
$n^{-3/2} (\alpha_s \sqrt{n})^l$ for $n\to \infty$. For the
$\MS$-scheme this is not the case and, as we will see, this scheme is
particularly inappropriate for large $n$.

\begin{figure}[h]
   \epsfxsize=11cm
   \centerline{\epsffile{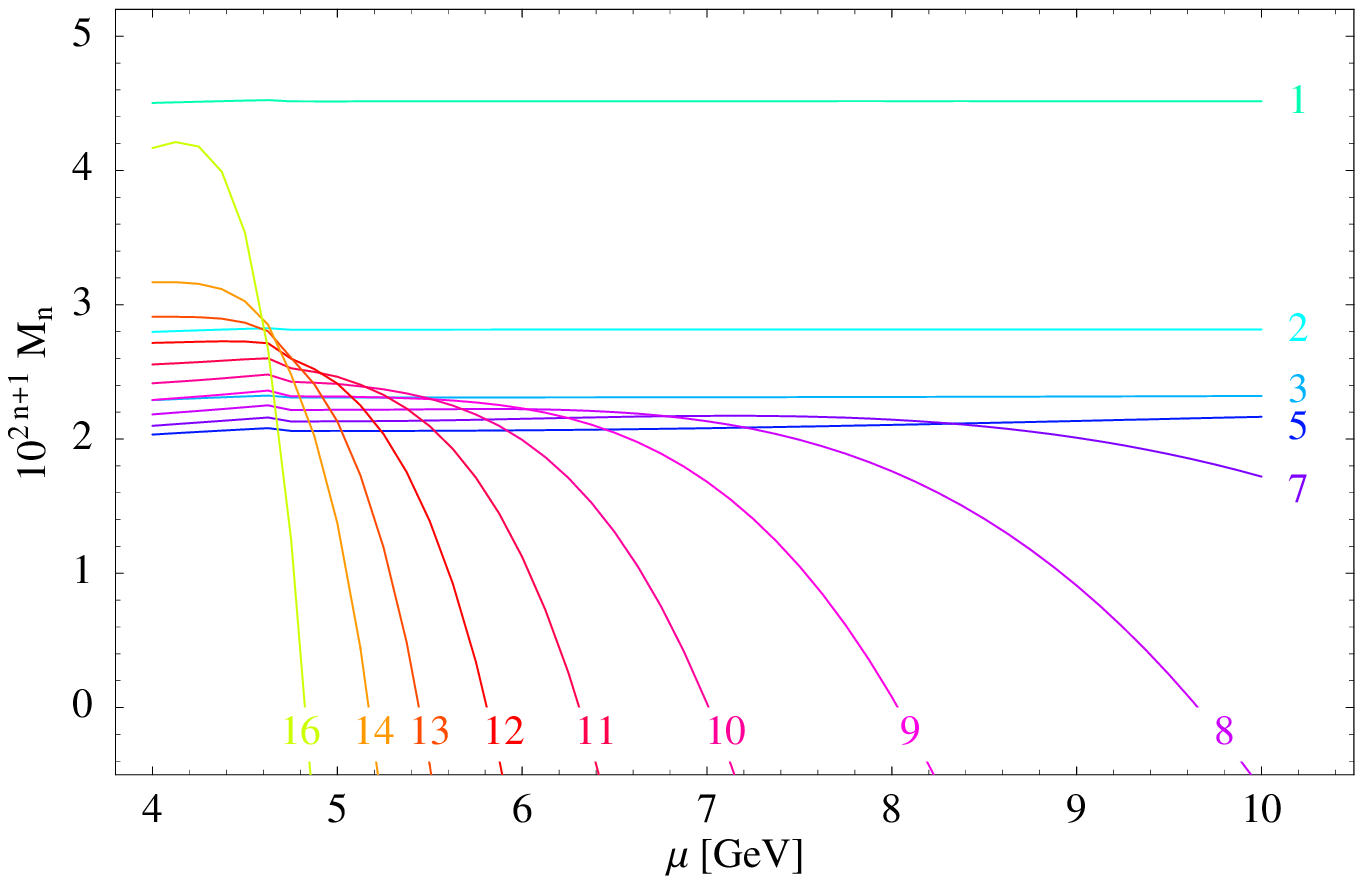} }
   \vspace*{0.2cm} \ccaption{}{Scale dependence of (some of) the first
 16 moments in the $\MS$-scheme. The moments are evaluated with
 $\overline{m}=4.184~{\rm GeV}$ in a fixed-order approach including
 terms up to $\cO(\alpha_s^3)$.
 \label{figMSfo}}
\end{figure}

In order to substantiate this point consider the scale dependence of
the first 16 moments evaluated in a fixed-order approach up to
$\cO(\alpha_s^3)$ in the $\MS$-scheme. The $n$-th moment has mass
dimensions $[m_b]^{-2n}$ and the moments in this paper are always
given in units $[{\rm GeV}]^{-2n}$. Since $C_{n,\MS}^{(30)}$ is not
known for $n>1$ we set $C_{n,\MS}^{(30)} = C_{1,\MS}^{(30)}$. We fix
$\overline{m}\equiv m_{\MS}(m_{\MS}) = 4.184$~GeV and then evaluate
$m_{\MS}(\mu)$ using the renormalization-group equations to four-loop
accuracy~\cite{Chetyrkin:2000yt}. We then use this value for the mass
and the scale $\mu$ to evaluate $M_{n, \MS}$ and vary the scale in the
region $4~{\rm GeV} \le \mu \le 10~{\rm GeV}$. As can be seen in
Figure~\ref{figMSfo}, for $n\le 6$ the moments are very stable for the
whole range of $\mu$, but for larger values of $n$ the scale
dependence deteriorates rapidly and it is clear that for $n\ge 8$ no
information can be extracted from these results any longer. The
precise shape of the curves in Figure~\ref{figMSfo} depends to some
extent on details as how to treat the flavour threshold and the value
of $C_{n,\MS}^{(30)}$, but the main point is not affected by these
issues.

\begin{figure}[h]
   \epsfxsize=11cm
   \centerline{\epsffile{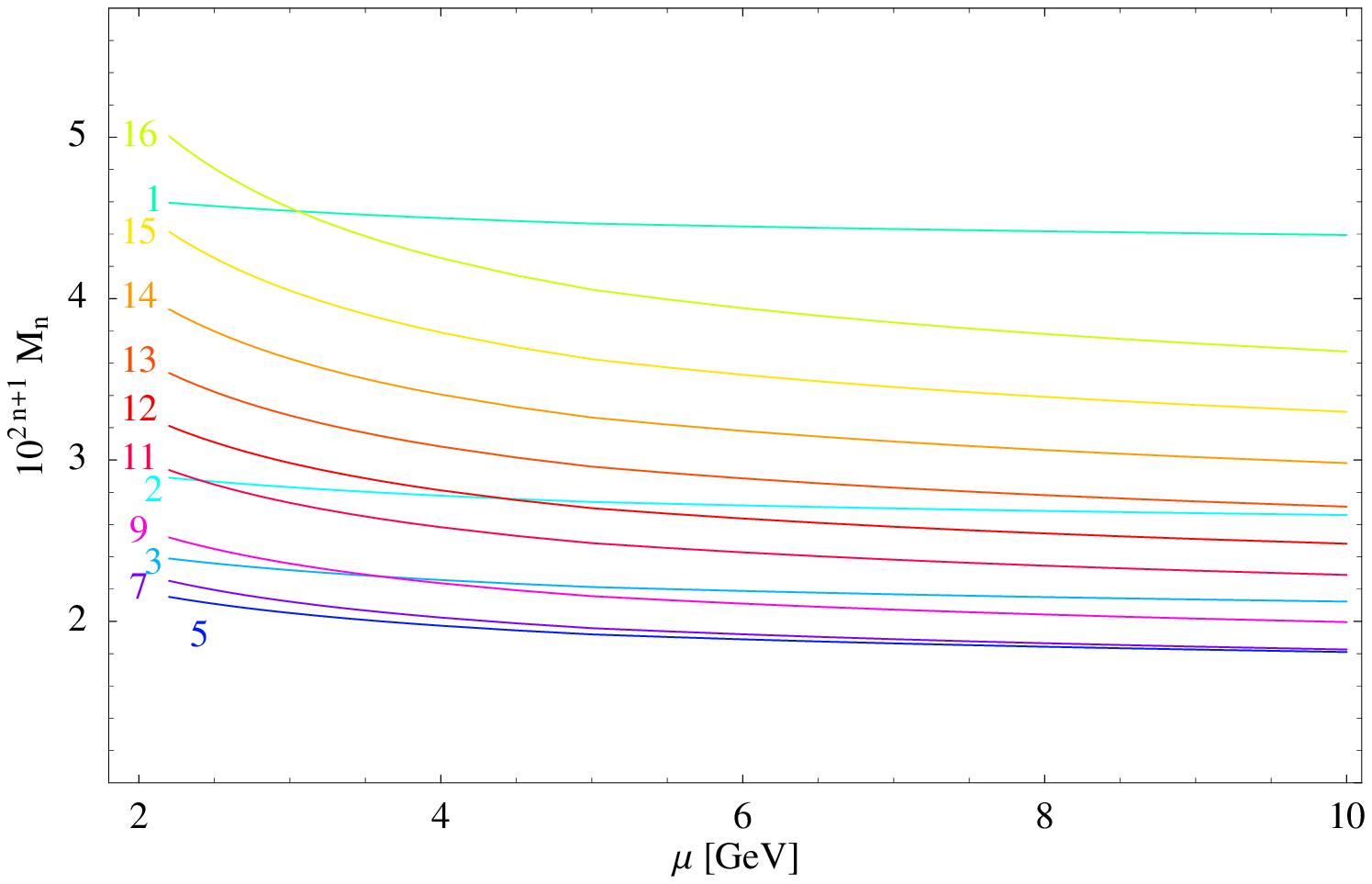} }
   \vspace*{0.2cm} \ccaption{}{Scale dependence of (some of) the first
 16 moments in the $\PS$-scheme with $m_\PS = 4.505$~GeV. The moments
 are evaluated in a fixed-order approach including terms up to
 $\cO(\alpha_s^3)$.
 \label{figPSfo}}
\end{figure}

The same exercise can be repeated for a threshold mass. As discussed
above in this case we would expect a somewhat better behaviour for
large $n$. To investigate this, we use the
$\PS$-scheme~\cite{Beneke:1998rk} and set $m_{\PS}\equiv
m_{\PS}(\mu_F=2~{\rm GeV}) = 4.505~{\rm GeV}$ which, using three-loop
conversion, corresponds to $\overline{m} = 4.184$~GeV. Since $\delta
m_{\PS}^{(1)} =C_F\, \mu_F/m_{\PS}$ we have to choose the
factorization scale $\mu_F \sim m_b\, \alpha_s$ to ensure the
additional suppression $\delta m_{X}^{(l)} \sim \alpha_s$ mentioned
above and the standard choice is $\mu_F=2~{\rm GeV}$. Again we
evaluate the first 16 moments varying the scale in the region
$2.5~{\rm GeV} \le \mu \le 20~{\rm GeV}$. The results depicted in
Figure~\ref{figPSfo} show a remarkable stability with respect to the
scale variation. The scale dependence does increase for increasing $n$
but remains in much better control than in the $\MS$-scheme.

Even though the results shown in Figures~\ref{figMSfo} and
\ref{figPSfo} nicely confirm our expectations it is clear that even in
the $\PS$-scheme the results become unreliable for large $n$. To
obtain a more complete picture we now turn to the evaluation of the
moments in the non-relativistic effective-theory approach and compare
these results with the fixed-order results of this section.

\section{Effective theory results, large $n$ \label{sec:III}}

As mentioned in the introduction, in the effective-theory approach we
start from the Schr\"odinger equation describing a non-relativistic
heavy quark pair with energy $E = \sqrt{s}-2m_b$ interacting through
the potential $-C_F\alpha_s/r$.  The cross section $R(s)$ is related
to the imaginary part of the corresponding Green function at the
origin. Working in dimensional regularization in $d=4-2\epsilon$
dimensions and minimally subtracting the $1/\epsilon$ ultraviolet
singularity, the leading order Coulomb Green function at the origin is
given by~\cite{Beneke:1999qg}
\begin{equation}
\label{Greenorig}
G^{(0)}_c(0,0;E) = - \frac{\alpha_s\, C_F\, m_b^2}{4\pi} 
\left( \frac{1}{2\lambda} + \frac{1}{2} \log \frac{-4 m_b E}{\mu^2}
       - \frac{1}{2} + \gamma_E + \psi(1-\lambda) \right)
\end{equation}
where $\lambda\equiv C_F\, \alpha_s/(2 \sqrt{-E/m_b})$ and the
leading-order cross section is given by
\begin{equation}
R(E) = 6 \pi N_c\, \frac{e_b^2}{m_b^2}\, 
{\rm Im}\left[ G_c(0,0;E)\right]
\label{Rlo}
\end{equation}
Higher-order corrections are computed by perturbative insertions of
higher-order corrections to the potential.  The moments are then
evaluated preforming the integration indicated in \Eqn{MnDef}. In the
literature different options on how to treat the prefactor $1/s^{n+1}$
have been used. Either, this factor can be expanded, writing
\begin{equation}
M_n = \int_{-\infty}^\infty
 \frac{2\, dE}{(2 m_b)^{2n+1}}\, e^{-\frac{n E}{m_b}} 
\left(1 - \frac{E}{2 m_b} + \frac{n E^2}{(2 m_b)^2}  
  +\ldots \right) R(E)
\label{MomExp}
\end{equation}
where the ellipses stand for higher-order terms in the
non-relativistic expansion, or it can be left unexpanded
\begin{equation}
M_n = \int_{-\infty}^\infty \frac{2\, dE}{(E+2 m_b)^{2n+1}} \, R(E)
\label{MomStd}
\end{equation}
\Eqns{MomExp}{MomStd} agree at NNLO in the effective theory, but will
differ considerably for small $n$. In this section we will use the
strictly expanded approach, \Eqn{MomExp}, but we will come back to
this issue in Section~\ref{sec:IV}.

\begin{figure}[h]
   \epsfxsize=11cm
   \centerline{\epsffile{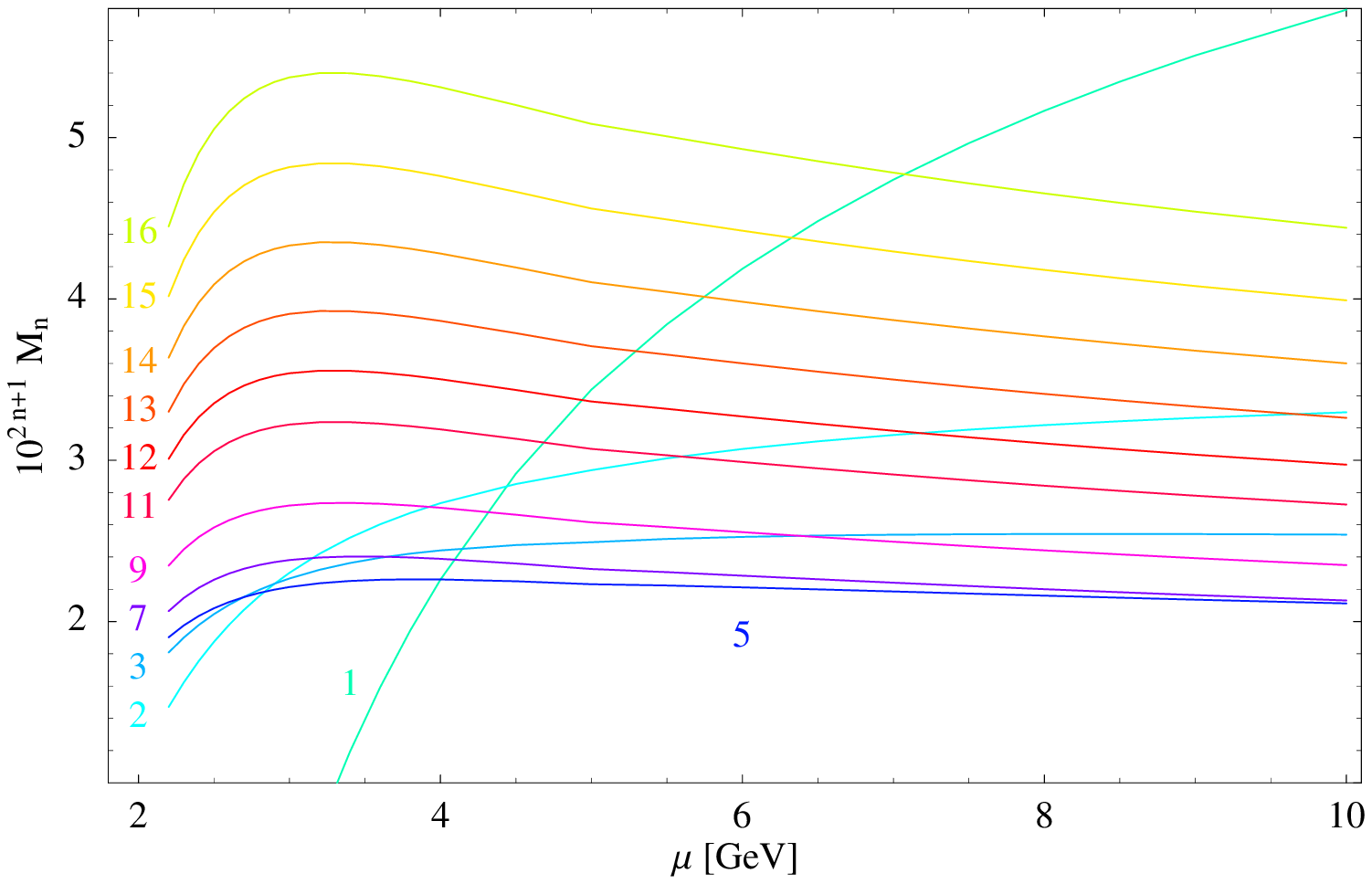} }
   \vspace*{0.2cm} \ccaption{}{Scale dependence of (some of) the first
 16 moments in the $\PS$-scheme with $m_\PS= 4.505$~GeV. The moments
 are evaluated in an effective-theory approach including terms up to
 NNLL accuracy.
 \label{figPSet}}
\end{figure}

In Figure~\ref{figPSet} we show the scale dependence of (some of) the
first 16 moments evaluated in the effective theory with $m_\PS=
4.505$~GeV. These results are complete at NNLO. Thus, counting
$\alpha_s \sim 1/\sqrt{n}$ they include all terms scaling like
$n^{-3/2} \left(\alpha_s \sqrt{n}\right)^l\, \alpha_s^2$. Furthermore
large logarithms are resummed counting $\alpha_s \log n \sim 1$. The
results presented here are complete at next-to-leading logarithmic
(NLL) accuracy and contain some known contributions at
next-to-next-to-leading logarithmic (NNLL)
accuracy~\cite{Pineda:2006gx}. The scale dependence of the first few
moments is very strong, indicating the expected breakdown of the ET
approach for small $n$. However, for $n\gtrsim 5$ the results are
very stable. Increasing $n$ further to $n\gtrsim 14$ leads to an
enhanced scale dependence. This is not unexpected, since missing
higher-order and NNLL terms as well as non-perturbative corrections
become increasingly important.

Of course, the scale dependence is at best a very rough indicator of
the reliability of the results.  In the following section we will
combine the results of Sections~\ref{sec:II} and \ref{sec:III}, including
all terms ${\cal O}(\alpha_s^3)$ of the FO approach  and
all (known) NNLL terms of the ET approach and investigate the relative
importance of the various corrections.

\section{Combined analysis \label{sec:IV}}

In this section we present the results of a combined approach,
i.e. results that are complete at ${\cal O}(\alpha_s^3)$ in the FO
approach and complete at NNLL in the ET approach. The results are
obtained simply by adding the FO and ET results and subtracting the
doubly counted terms. In order to obtain the doubly counted terms we
expand the ET result in the coupling $\alpha_s$ and retain all terms
of ${\cal O}(\alpha_s^3)$. Note that these terms depend on the precise
implementation of the non-relativistic expansion. In particular, they
depend on whether \Eqn{MomExp} or \Eqn{MomStd} is used. Any
implementation that is equivalent at large $n$ of the ET result can be
used, as long as the subtraction terms are treated consistently. 

Using the implementation according to \Eqn{MomExp} and expanding the
ET result in $\alpha_s$ leads to integrals of the form
\begin{equation}
\int_{-\infty}^{\infty} \frac{2\, dE}{(2 m_b)^{2n+1}}\, 
e^{-\frac{n E}{m_b}}\
{\rm Im}\left[ \left(\frac{m_b}{-E}\right)^x \, 
       \log^k\left(\frac{-E\, m_b}{\mu^2}\right) \right]
\label{intA}
\end{equation}
for the corresponding contribution to $M_n$, where $E=E+i 0^+$ is
understood. These integrals can either be computed numerically or
obtained analytically by differentiation with respect to $y$ of
\begin{eqnarray}
\cI_{\rm exp}(n,x,y) &\equiv& 
\int_{-\infty}^{\infty} \frac{2\, dE}{(2 m_b)^{2n+1}}\, 
e^{-\frac{n E}{m_b}}\ 
{\rm Im}\left[ \left(\frac{m_b}{-E}\right)^x \, 
       \left(\frac{-E\, m_b}{\mu^2}\right)^y \right]
\label{IntExp} \\
&=&  \frac{\pi}{(2 m_b)^{2n}}
\left(\frac{m_b^2}{\mu^2}\right)^y 
\frac{n^{-1+x-y}}{\Gamma(x-y)}
\nonumber
\end{eqnarray}
The corresponding integral using the implementation according to
\Eqn{MomStd} is given by
\begin{eqnarray}
\cI_{\rm std}(n,x,y) &\equiv & 
\int_{-\infty}^{\infty} \frac{2\, dE}{(E+2 m_b)^{2n+1}}\, 
{\rm Im}\left[ \left(\frac{m_b}{-E}\right)^x \, 
       \left(\frac{-E\, m_b}{\mu^2}\right)^y \right]
\label{IntStd}\\
&=& \frac{\pi}{(2 m_b)^{2n}} \left(\frac{m_b^2 }{\mu^2}\right)^y
\frac{2^{1-x+y}\, \Gamma(x-y+2n)}{\Gamma(2n+1)\Gamma(x-y)} \nonumber
\end{eqnarray}
In the derivation of these results we assumed that $\mu$ is
independent of $E$.

\begin{table}[h]
\begin{center}
\bigskip
\begin{tabular}{|l|c|c|c|c|c|c|c|}
\hline
$n$ & 1 & 2 & 3 & 4 & 5 & 6 & 7  \\
\hline
`exact LO' & 9.235 & 5.021 & 3.998 & 3.701 & 3.713 & 3.914 & 4.267 \\
${\cal O}(\alpha_s^0)$ 
& 5.458 & 2.377 & 1.594 & 1.275 & 1.124 & 1.053 & 1.030    \\
${\cal O}(\alpha_s^1)$ 
& 8.294 & 4.124 & 3.028 & 2.600 & 2.430 & 2.394 & 2.445  \\
${\cal O}(\alpha_s^2)$ 
& 9.065 & 4.795 & 3.704 & 3.321 & 3.224 & 3.287 & 3.463  \\
${\cal O}(\alpha_s^3)$ 
& 9.211 & 4.976 & 3.926 & 3.595 & 3.561 & 3.702 & 3.975  \\
${\cal O}(\alpha_s^4)$ 
& 9.233 & 5.014 & 3.983 & 3.676 & 3.673 & 3.853 & 4.176  \\
${\cal O}(\alpha_s^5)$ 
& 9.236 & 5.020 & 3.995 & 3.696 & 3.704 & 3.899 & 4.242  \\ 
\hline
\end{tabular}
\end{center}
\ccaption{}{Comparison of the exact LO result in the effective theory
  to the expanded results, \Eqn{expandLOint}. All entries are
  multiplied by $10^{2n+1}$.
\label{tab:ETexpand} }
\end{table}

In order to illustrate the procedure, let us take the leading-order
result in the effective theory, given in \Eqns{Greenorig}{Rlo} and
expand it in $\alpha_s$ to say  ${\cal O}(\alpha_s^5)$
\begin{eqnarray}
\label{expandLO}
R(E) &=& \frac{3}{4} N_c\, C_F\, e_b^2\, \Bigg( 
  - \frac{1}{\ell} 
  + \alpha_s \left(1- \log \frac{-4 m_b E}{\mu^2}\right)
  + 2\, \alpha^2_s\, \ell \, \psi^{(1)}(1)
\\
 &&\qquad 
  -\ \alpha^3_s\, \ell^2\,  \psi^{(2)}(1)
  + \frac{1}{3} \alpha^4_s\, \ell^3\,  \psi^{(3)}(1)
  - \frac{1}{12} \alpha^5_s\, \ell^4\,  \psi^{(4)}(1)
  + \ldots \Bigg)
\nonumber
\end{eqnarray}
where we introduced $\ell\equiv \lambda/\alpha_s = C_F/(2
\sqrt{-E/m_b})$ and $\psi^{(k)}$ denotes the $k$-th derivative of the
$\psi$-function. Integrating this series according to
\Eqn{MomExp}, dropping the higher-order terms in $E/m_b$, we get
\begin{eqnarray}
\label{expandLOint}
M^{(0)}_n &=&  \frac{3\, n^{-3/2}}{4 (2m_b)^{2n}} N_c\, e_b^2\, 
\Bigg( \sqrt{\pi} + \bar{\alpha}\, \pi 
+ \bar{\alpha}^2\, \sqrt{\pi}  \, \psi^{(1)}(1)
\\
&& \qquad -\ \bar{\alpha}^3 \, \frac{\pi}{4}\, \psi^{(2)}(1)
 + \bar{\alpha}^4 \, \frac{\sqrt{\pi}}{12}\, \psi^{(3)}(1)
 - \bar{\alpha}^5\, \frac{\pi}{192}\, \psi^{(4)}(1) + \ldots \Bigg)
\nonumber
\end{eqnarray}
with $\bar{\alpha} \equiv C_F (\alpha_s\sqrt{n})$. We can now check
how the expanded result, \Eqn{expandLOint} approaches the `exact'
leading-order result in the effective theory. This is done in
Table~\ref{tab:ETexpand}, where we show the results of performing
according to \Eqn{MomExp} the integration of $R(E)$ as given in
\Eqn{Rlo}. As in the derivation of \Eqn{expandLOint} we drop
higher-order terms in $E/m_b$ and, for convenience, multiply by
$10^{2n+1}$. The results for $n\le 7$ with $m_b=4.505~{\rm GeV}$ and
$\mu=4.5~{\rm GeV}$ are shown in the second row, labelled `exact
LO'. The other rows contain the successive approximations given in
\Eqn{expandLOint}, with $\alpha_s\equiv \alpha_s(\mu=4.5~{\rm GeV}) =
0.2198$. As expected, the expanded results approach the `exact result'
faster for smaller values on $n$. Including all terms up to ${\cal
O}(\alpha_s^3)$ the relative error is $\{0.3\%, 0.9\%, 1.8\%, 3.0\%,
4.3\%, 5.7\%,7.4\%\}$ for $n=\{1,2,3,4,5,6,7\}$ respectively. The
doubly counted terms to be subtracted in the combined analysis at this
order correspond to the terms given in the row labelled ${\cal
O}(\alpha_s^3)$.

\begin{figure}[h]
   \epsfxsize=11cm
   \centerline{\epsffile{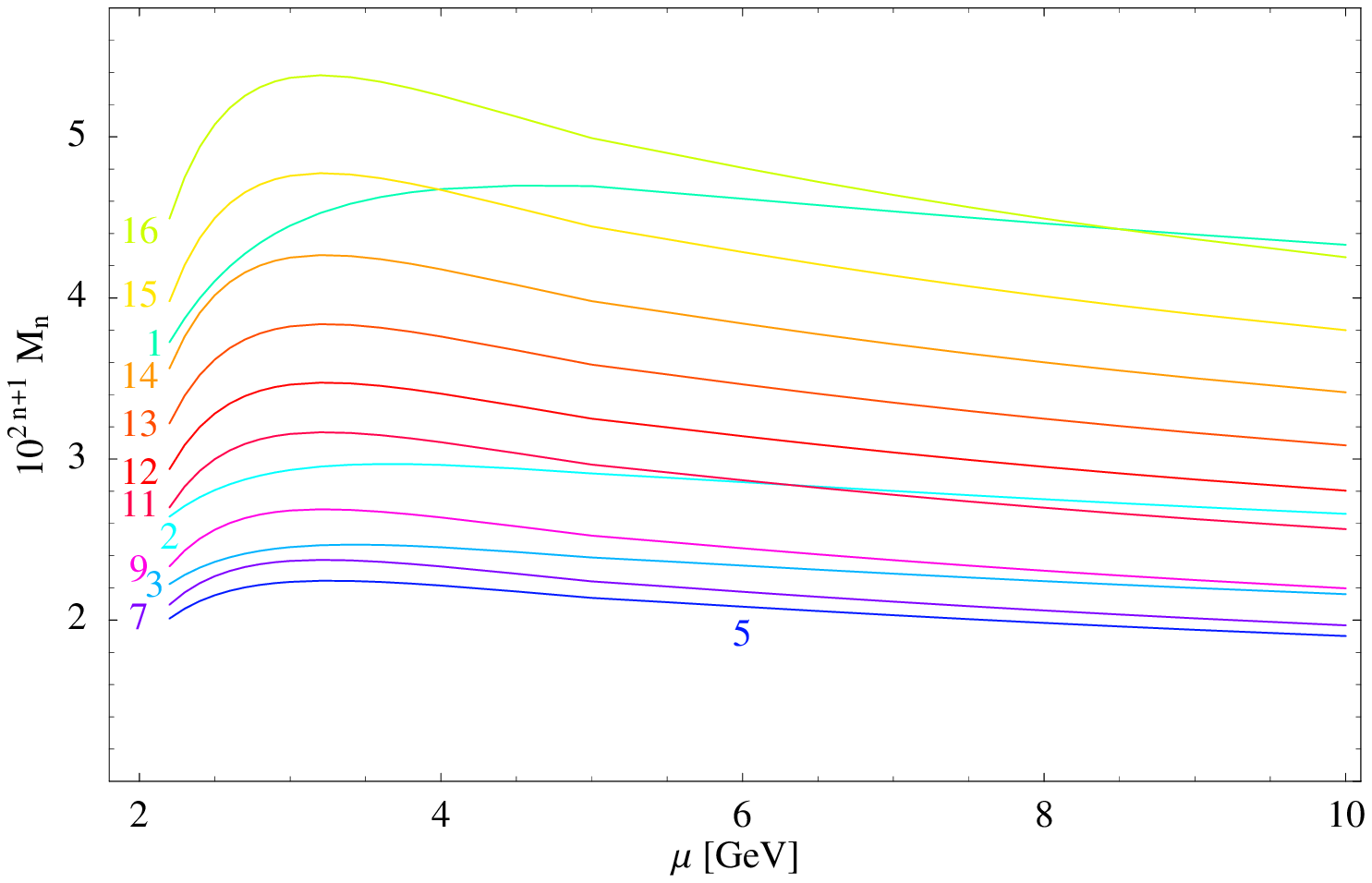} }
   \vspace*{0.2cm} \ccaption{}{Scale dependence of the first 16
 moments in the $\PS$-scheme with $m_\PS = 4.505$~GeV. The moments are
 evaluated in a combined approach including terms up to NNLL accuracy
 and order $\cO(\alpha_s^3)$.
 \label{figPScb}}
 \end{figure}

\begin{table}[h]
\begin{center}
\bigskip
\begin{tabular}{|l|c|c|c|c|c|c|c|c|c|}
\hline
$n$ & 1 & 2 & 3 & 4 & 6 & 8 & 10 & 13 & 16 \\
\hline
$10^{2n+1} M_n$ & 4.70 & 2.94 & 2.42 & 2.23 & 2.21 & 2.42 & 2.79 &
3.68 & 5.13 \\
comb/ET & 1.61 & 1.03 & 0.98 & 0.97 & 0.97 & 0.97 & 0.97 & 0.97 & 0.99\\ 
comb/FO& 1.05 & 1.07 & 1.09 & 1.10 & 1.14 & 1.16 & 1.19 & 1.22 & 1.24  \\
\hline
\end{tabular}
\end{center}
\ccaption{}{Comparison of the combined evaluation of selected moments
$M_n$ with the ET and FO approach. The moments are evaluated with
$m_\PS=4.505~{\rm GeV}$ with the scale $\mu = 4.5~{\rm GeV}$.
\label{tab:cbetfo} }
\end{table}

\begin{figure}[h]
   \epsfxsize=11cm
   \centerline{\epsffile{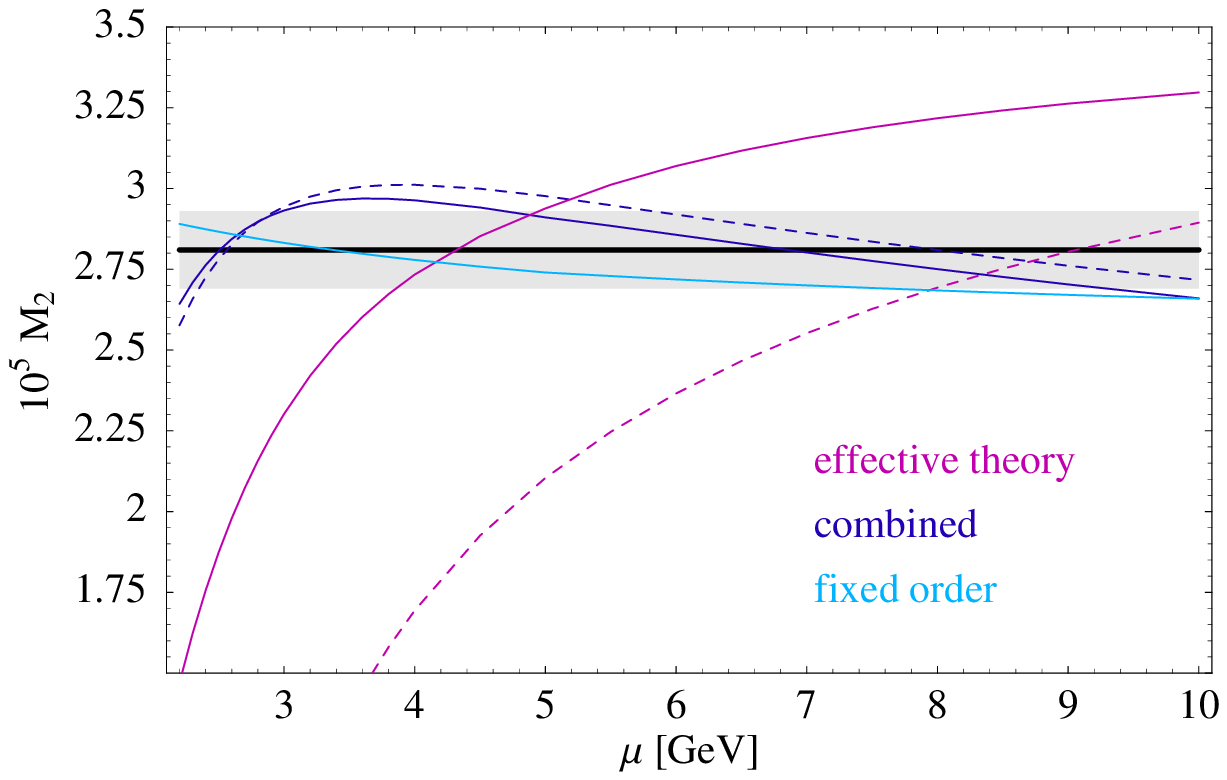} }
   \vspace*{0.2cm} \ccaption{}{Scale dependence of the second moment
 in the $\PS$-scheme evaluated using a FO (light blue curve), ET
 (magenta curves) and a combined approach (dark blue curves). The
 dashed curves have been obtained using \Eqn{MomExp}, whereas the solid
 curves have been obtained using \Eqn{MomStd}.
 \label{PSmom2}}
 \end{figure}

Repeating this exercise with the full NNLL effective-theory result and
combining the FO and ET results using the definition \Eqn{MomExp} we
evaluate again the first 16 moments and depict their scale dependence
in Figure~\ref{figPScb}. Comparing Figures~\ref{figPScb} and
\ref{figPSet} we note that, as expected, the difference is small for
large $n$ and large for small $n$. The FO corrections to the ET
results are $\lesssim 5\%$ for $n=10$ (except for very small scales),
increasing to $\lesssim 10\%$ for $n=4$. For $n\leq 2$ the corrections
completely change the shape of the curve, indicating the importance of
relativistic corrections to the non-relativistic sum rules. On the
other hand, comparing Figures~\ref{figPScb} and \ref{figPSfo} the
situation is just reversed. For small $n$ the corrections are small
(except for very small scales) and they increase with increasing $n$
indicating the importance of resumming terms $\left(\alpha_s
\sqrt{n}\right)^l$ for large $n$. This is also confirmed by
Table~\ref{tab:cbetfo}, where we list the value of some combined
moments (second row), as well as the ratio of the combined moment to
the ET result (third row) and FO result (fourth row) respectively. We
should stress that the ratios in Table~\ref{tab:cbetfo} depend on the
scale choice $\mu=4.5~{\rm GeV}$ and only give an incomplete
picture. In particular, the corrections to the ET result for
$n\in\{2,3,4\}$ are larger than what might be inferred from
Table~\ref{tab:cbetfo}. This is illustrated in Figure~\ref{PSmom2}
where the scale dependence of the second moment is plotted and
compared to the experimental moment with its error, indicated by the
black line and the grey rectangle. The FO result is plotted as the
light blue line. The ET result is evaluated using \Eqn{MomExp} (solid
dark blue line) and \Eqn{MomStd} (dashed dark blue line). As mentioned
above, the two implementations differ considerably (for small
$n$). However, the corresponding combined results, depicted as solid
and dashed magenta lines respectively, are virtually independent of
the implementation, since differences in treating higher-order in $n$
terms are compensated for by adding the full $n$ dependence up to
${\cal O}(\alpha_s^3)$ through the FO result. From Figure~\ref{PSmom2}
we can also see that the value of 1.03 given in Table~\ref{tab:cbetfo}
for the ratio of the combined and ET result for the second moment is a
coincidence of the scale choice $\mu=4.5~{\rm GeV}$ and not
necessarily indicative of the typical size of the corrections.

The experimental moments used in Figure~\ref{PSmom2} and the following
plots have been determined by taking into account $M_n^{\rm res}$, the
contribution due to the six lowest resonances, and using perturbative
QCD in the region $\sqrt{s} > 11.2~{\rm GeV}$ to obtain the continuum
contribution $M_n^{\rm cont}$. This follows closely
Ref.~\cite{Kuhn:2001dm} from which we also adopt the treatment of
$M_n^{\rm lin}$, the additional contribution in the region $\sqrt{s} >
11.2~{\rm GeV}$. Due to the uncertainty and the lack of precise
experimental data in the region around and just above threshold we add
the errors linearly. The experimental moments and their errors are
listed in Table~\ref{tab:expmom}. The first four moments agree within
errors with those given in Ref.~\cite{Kuhn:2007vp}, but our
experimental moments have a larger error.

\begin{table}[h]
\begin{center}
\bigskip
\begin{tabular}{|l|c|c|c|c|c|c|c|c|c|}
\hline
$n$ & 1 & 2 & 3 & 4 & 6 & 8 & 10 & 13 & 16 \\
\hline
$10^{2n+1} M_n^{\rm exp}$ & 
4.51 & 2.81 & 2.31 & 2.13 & 2.11 & 2.30 & 2.64 & 3.40 & 4.53 \\
$10^{2n+1} \delta M_n^{\rm exp}$ & 
0.15 & 0.12 & 0.10 & 0.08 & 0.07 & 0.06 & 0.06 & 0.06 & 0.07   \\ 
\hline
\end{tabular}
\end{center}
\ccaption{}{Values of some selected experimental moments and their
errors.
\label{tab:expmom} }
\end{table}

\begin{figure}[h]
   \epsfxsize=11cm \centerline{\epsffile{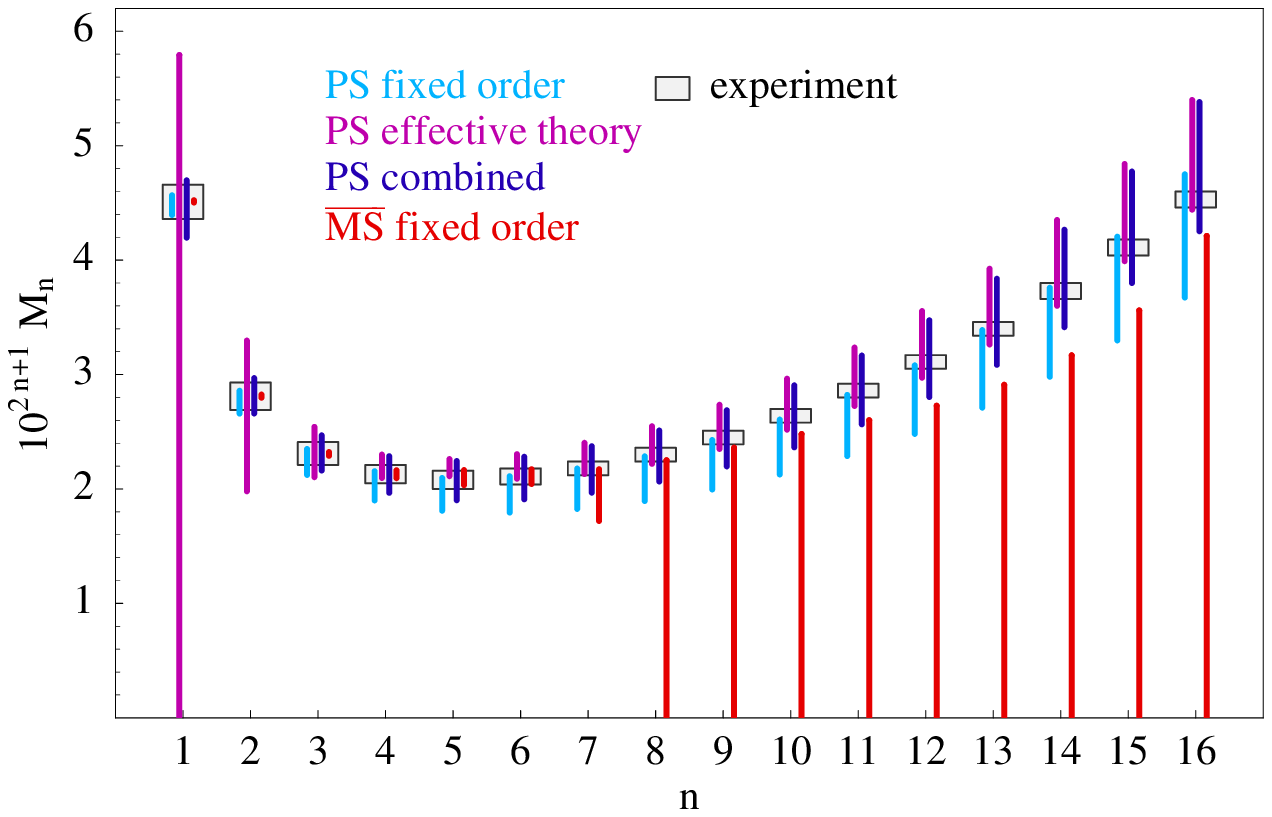} }
   \vspace*{0.2cm} \ccaption{}{Comparison of the experimental moments
     to the PS-scheme calculation in a FO approach (left/light blue
     bands), in an ET approach (middle/magenta bands) and in a
     combined approach (right/dark blue bands). The bands have been
     obtained by varying $2.5~{\rm GeV} \leq \mu \leq 10~{\rm
     GeV}$. Similar bands for the $\MS$-scheme are shown in red.
 \label{fig:rangePS}}
\end{figure}

In order to get a better understanding of the relative importance of
the various corrections and the range of applicability of the various
approximations we compare the first 16 moments in the PS-scheme to the
experimental moments. We evaluate the moments in the FO, ET and
combined approach as well as in the $\MS$ fixed-order approach. Note
that $\overline{m}=4.184~{\rm GeV}$ has been determined by requiring
the first moment in the $\MS$ fixed-order approach to agree with the
experimental value. This then fixes $m_{\PS} = 4.505~{\rm GeV}$ and
all further moments. For the PS-scheme we vary the scale in the range
$2.5~{\rm GeV} \leq \mu \leq 10~{\rm GeV}$, whereas for the $\MS$
scheme we vary the scale in the range $4~{\rm GeV} \leq \mu \leq
10~{\rm GeV}$ as explained in Section~\ref{sec:I}.  The results are
shown in Figure~\ref{fig:rangePS}. Note that the scale $\mu=4.5~{\rm
GeV}$ chosen in Table~\ref{tab:cbetfo} is close to the upper end of
the scale variation shown in Figure~\ref{fig:rangePS}. This explains
why the values in Table~\ref{tab:cbetfo} are generically larger than
the experimental moments given in Table~\ref{tab:expmom}.  The most
striking feature of Figure~\ref{fig:rangePS} is that in the PS-scheme
all three approaches give very similar results. In particular, the FO
approach gives good results even for $n=16$. The ET approach seems to
be valid down to $n=3$ and only breaks down for $n\lesssim 2$. We note
this seems to be a general feature of any suitably defined threshold
mass. In particular, we have checked that for the
RS-mass~\cite{Pineda:2001zq} the results are very similar. On the
other hand, the situation is rather different in the $\MS$ scheme. The
FO approach gives excellent results for $n\lesssim 7$ and then breaks
down abruptly. This can also be inferred from Figure~\ref{figMSfo}.
Had we chosen to limit the scale variation by say $\mu < 7~{\rm GeV}$
we would have obtained good results up to $n=9$.

\begin{figure}[h]
   \epsfxsize=11cm
   \centerline{\epsffile{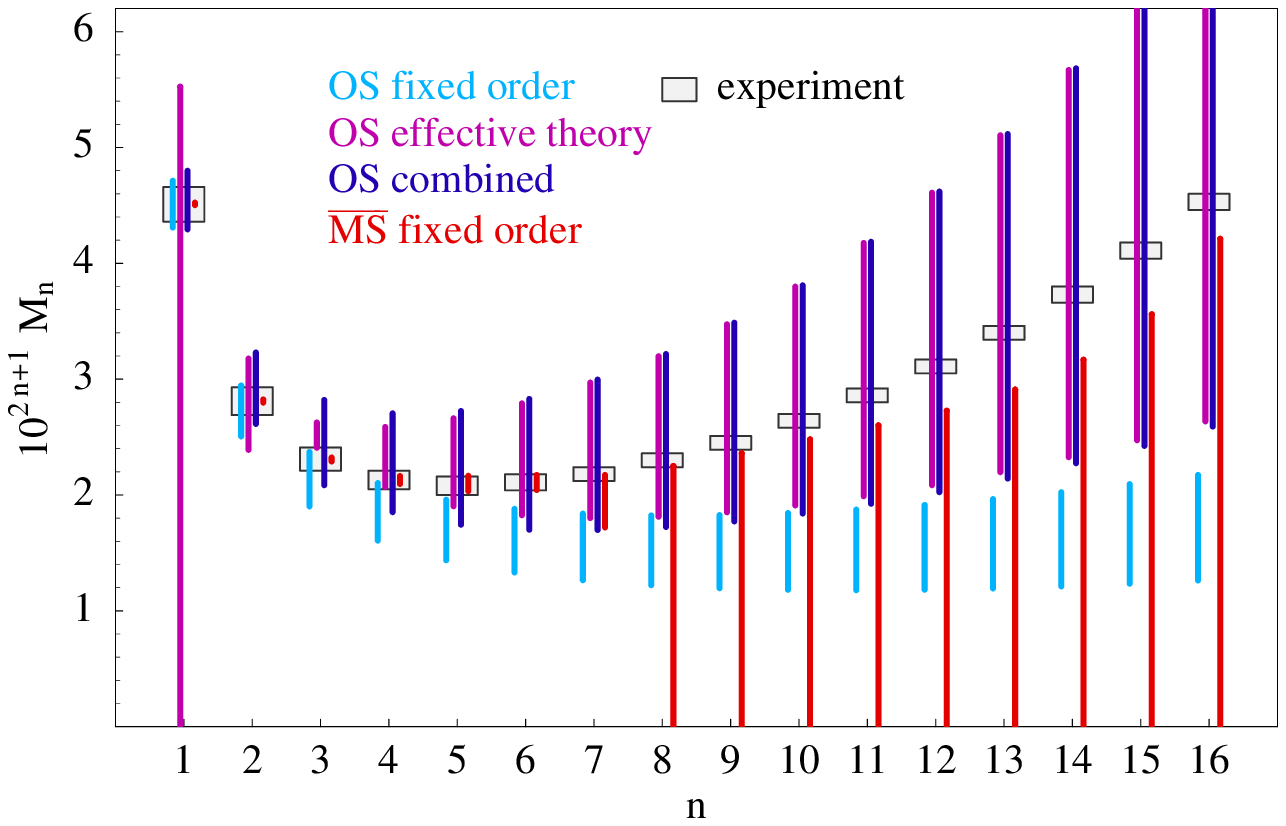} }
   \vspace*{0.2cm} \ccaption{}{Comparison of the experimental moments
     to the moments computed in the OS scheme with the pole mass set
     to $m=4.85~{\rm GeV}$ in a FO approach (left/light blue bands),
     in an ET approach (middle/magenta bands) and in a combined
     approach (right/dark blue bands). The bands have been obtained by
     varying $2.5~{\rm GeV} \leq \mu \leq 10~{\rm GeV}$. Similar bands
     for the $\MS$-scheme are shown in red.
 \label{fig:rangeOS}}
\end{figure}

Finally, we present a similar plot for the OS scheme.  We fix the
value of the pole mass to make the first moment in the FO approach to
agree with the experimental moment. This results in $m=4.85~{\rm
GeV}$. The we proceed as in the case of the PS scheme. As mentioned at
the beginning, the OS scheme is not well suited for a precise
determination of quark masses and we would expect the results to be
less consistent than with other mass definitions. This is what we find
in Figure~\ref{fig:rangeOS}. The FO results are inconsistent with the
experimental values of the moments for $n\ge 6$. Accordingly, the
corrections $\left(\alpha_s \sqrt{n}\right)^l$ are more important than
in the PS scheme and bring the combined results into agreement with
the ET results and the experimental values. The ET results agree with
the experimental moments for all values of $n$, but for $n=1$ the
scale dependence is enormous, making the result meaningless. Overall,
the scale dependence is considerably larger than in the PS scheme, in
agreement with our expectations.

\section{Conclusions and outlook \label{sec:C}}

The main result of this analysis is that there is no need to make the
standard separation into large-$n$ and small-$n$ analyses of the
$b\bar{b}$ sum rules. For a suitably defined threshold mass, the
fixed-order results are remarkably consistent even for large values of
$n$. With hindsight one might argue that from a numerical point of
view the expected breakdown does not happen at $\alpha_s \sqrt{n} \sim
1$ but rather at $(\alpha_s/\pi)\, \sqrt{n} \sim 1$. Even in the case
of the $\MS$-mass where the large-$n$ behaviour is worse due to the
presence of terms $\left(\alpha_s\, n\right)^l$, a fixed-order
approach is applicable for values of $n$ up to $n\simeq 6$. On the
other hand the non-relativistic sum rule can also be applied for
values of $n$ that are much smaller than what naively could have been
expected. Overall, we obtain a very consistent picture. With the
availability of the NNNLO corrections in the effective
theory~\cite{ETnnnlo} and the prospect of complete results in the
fixed-order approach at ${\cal O}(\alpha_s^3)$ also for $n>1$, the sum
rule is likely to be the observable of choice for bottom quark mass
determinations. Due to the large range of $n$ that can be used,
non-perturbative corrections are well under control and additional
effects such as non-vanishing charm mass~\cite{masseff,
Corcella:2002uu} can be included as well. In view of the progress on
the theoretical side, more precise experimental data of the $b$-quark
cross section just above threshold would be most welcome.

\vspace*{0.5em}
\noindent
\subsubsection*{Acknowledgement}
This work is supported in part by the European Community's Marie-Curie
Research Training Network under contract MRTN-CT-2006-035505 `Tools
and Precision Calculations for Physics Discoveries at Colliders'.



\begin{thebibliography}{99}

\bibitem{Novikov:1976tn}
V.~A.~Novikov {\it et al.},
Phys.\ Rev.\ Lett.\  {\bf 38}, 626 (1977)
[Erratum-ibid.\  {\bf 38}, 791 (1977)].

\bibitem{Chetyrkin:1997mb}
K.~G.~Chetyrkin, J.~H.~Kuhn and M.~Steinhauser,
Nucl.\ Phys.\  B {\bf 505}, 40 (1997)
[arXiv:hep-ph/9705254].

\bibitem{Boughezal:2006uu}
R.~Boughezal, M.~Czakon and T.~Schutzmeier,
Nucl.\ Phys.\ Proc.\ Suppl.\  {\bf 160}, 160 (2006)
[arXiv:hep-ph/0607141].

\bibitem{Chetyrkin:2006xg}
K.~G.~Chetyrkin, J.~H.~Kuhn and C.~Sturm,
Eur.\ Phys.\ J.\  C {\bf 48}, 107 (2006)
[arXiv:hep-ph/0604234].

\bibitem{Boughezal:2006px}
R.~Boughezal, M.~Czakon and T.~Schutzmeier,
Phys.\ Rev.\  D {\bf 74}, 074006 (2006)
[arXiv:hep-ph/0605023].

\bibitem{Kuhn:2007vp}
J.~H.~Kuhn, M.~Steinhauser and C.~Sturm,
arXiv:hep-ph/0702103.

\bibitem{Kuhn:2001dm}
J.~H.~Kuhn and M.~Steinhauser,
Nucl.\ Phys.\  B {\bf 619}, 588 (2001)
[Erratum-ibid.\  B {\bf 640}, 415 (2002)]
[arXiv:hep-ph/0109084].

\bibitem{Corcella:2002uu}
G.~Corcella and A.~H.~Hoang,
Phys.\ Lett.\  B {\bf 554}, 133 (2003)
[arXiv:hep-ph/0212297].

\bibitem{Brambilla:2004jw}
N.~Brambilla, A.~Pineda, J.~Soto and A.~Vairo,
Rev.\ Mod.\ Phys.\  {\bf 77}, 1423 (2005).

\bibitem{BmassNR}
K.~Melnikov and A.~Yelkhovsky,
Phys.\ Rev.\  D {\bf 59}, 114009 (1999)
[arXiv:hep-ph/9805270]; \\
A.~A.~Penin and A.~A.~Pivovarov,
Nucl.\ Phys.\ B {\bf 549}, 217 (1999)
[arXiv:hep-ph/9807421]; \\
A.~H.~Hoang,
Phys.\ Rev.\ D {\bf 61}, 034005 (2000)
[arXiv:hep-ph/9905550]; \\
M.~Beneke and A.~Signer,
Phys.\ Lett.\ B {\bf 471} (1999) 233
[arXiv:hep-ph/9906475].


\bibitem{resumLog}
 A.~H.~Hoang, A.~V.~Manohar, I.~W.~Stewart and T.~Teubner,
Phys.\ Rev.\ Lett.\  {\bf 86}, 1951 (2001)
[arXiv:hep-ph/0011254]; \\
A.~Pineda,
Phys.\ Rev.\ D {\bf 66}, 054022 (2002) 
[arXiv:hep-ph/0110216]; \\
A.~Pineda,
Phys.\ Rev.\ D {\bf 65}, 074007 (2002) 
[arXiv:hep-ph/0109117]; \\
A.~H.~Hoang,
Phys.\ Rev.\ D {\bf 69}, 034009 (2004) 
[arXiv:hep-ph/0307376]; \\
A.~Pineda and A.~Signer,
Nucl.\ Phys.\  B {\bf 762}, 67 (2007)
[arXiv:hep-ph/0607239].


\bibitem{Pineda:2006gx}
A.~Pineda and A.~Signer,
Phys.\ Rev.\  D {\bf 73}, 111501 (2006)
[arXiv:hep-ph/0601185].

\bibitem{massdef}
I.~I.~Y.~Bigi, M.~A.~Shifman and N.~Uraltsev,
Ann.\ Rev.\ Nucl.\ Part.\ Sci.\  {\bf 47}, 591 (1997)
[arXiv:hep-ph/9703290]; \\
A.~H.~Hoang, Z.~Ligeti and A.~V.~Manohar,
Phys.\ Rev.\ Lett.\  {\bf 82}, 277 (1999)
[arXiv:hep-ph/9809423].

\bibitem{Beneke:1998rk}
M.~Beneke,
Phys.\ Lett.\ B {\bf 434}, 115 (1998)
[arXiv:hep-ph/9804241].

\bibitem{Pineda:2001zq}
A.~Pineda,
JHEP {\bf 0106}, 022 (2001)
[arXiv:hep-ph/0105008].

\bibitem{Shifman:1978bx}
M.~A.~Shifman, A.~I.~Vainshtein and V.~I.~Zakharov,
Nucl.\ Phys.\  B {\bf 147}, 385 (1979).

\bibitem{Broadhurst:1994qj}
D.~J.~Broadhurst, P.~A.~Baikov, V.~A.~Ilyin, J.~Fleischer,
O.~V.~Tarasov and V.~A.~Smirnov, 
Phys.\ Lett.\  B {\bf 329}, 103 (1994)
[arXiv:hep-ph/9403274].


\bibitem{Chetyrkin:2000yt}
K.~G.~Chetyrkin, J.~H.~Kuhn and M.~Steinhauser,
Comput.\ Phys.\ Commun.\  {\bf 133}, 43 (2000)
[arXiv:hep-ph/0004189].

\bibitem{Beneke:1999qg}
M.~Beneke, A.~Signer and V.~A.~Smirnov,
Phys.\ Lett.\ B {\bf 454}, 137 (1999)
[arXiv:hep-ph/9903260]; \\
M.~Beneke, 
in: Proceedings of the 8th International Symposium on Heavy 
Flavor Physics (Heavy Flavors 8), Southampton, England, 25-29 Jul
1999, [arXiv:hep-ph/9911490].

\bibitem{ETnnnlo}
M.~Beneke, Y.~Kiyo and K.~Schuller,
Nucl.\ Phys.\  B {\bf 714}, 67 (2005)
[arXiv:hep-ph/0501289]; \\
M.~Beneke, Y.~Kiyo and K.~Schuller,
arXiv:0705.4518 [hep-ph]; \\
M.~Beneke, Y.~Kiyo and A.~A.~Penin,
arXiv:0706.2733 [hep-ph].

\bibitem{masseff}
A.~H.~Hoang, J.~H.~Kuhn and T.~Teubner,
Nucl.\ Phys.\  B {\bf 452}, 173 (1995)
[arXiv:hep-ph/9505262]; \\
K.~G.~Chetyrkin, A.~H.~Hoang, J.~H.~Kuhn, M.~Steinhauser and T.~Teubner,
Eur.\ Phys.\ J.\  C {\bf 2}, 137 (1998)
[arXiv:hep-ph/9711327].



\end{thebibliography}
\end{document}